\documentclass[12pt,preprint]{aastex}
\usepackage{graphicx}
\slugcomment{Accepted by AJ}

\shorttitle{FUV Spectroscopy of Nova-Like Variables
}
\shortauthors{}

\begin{document}

\title{Far-Ultraviolet Spectroscopy of the Nova-Like Variable KQ~Monocerotis: A
New SW~Sextantis Star?}


\author{Aaron Wolfe \& Edward M. Sion
}
\affil{Astronomy \& Astrophysics, Villanova University, \\
800 Lancaster Avenue, Villanova, PA 19085, USA}
\email{awolfe03@villanova.edu, edward.sion@villanova.edu}

\author{Howard E. Bond} 
\affil{Department of Astronomy \& Astrophysics, Pennsylvania State University,
University Park, PA 16802; Space Telescope Science Institute, 3700 San Martin Dr., Baltimore, MD 21218
Current address: 9615 Labrador Ln., Cockeysville, MD 21030}

\email{bond@stsci.edu}

\begin{abstract}

New optical spectra obtained with the SMARTS 1.5m telescope and archival {\it
IUE\/} far-ultraviolet spectra of the nova-like variable KQ~Mon are discussed.
The optical spectra reveal Balmer lines in absorption as well as He I absorption superposed on a blue continuum. The 2011 optical spectrum is similar to the KPNO
2.1m IIDS spectrum we obtained 33 years earlier except that the Balmer and He I absorption is strogner in 2011. Far-ultraviolet {\it IUE\/} spectra
reveal deep absorption lines due to C II, Si III, Si IV, C IV, and He II, but no
P~Cygni profiles indicative of wind outflow. We present the results of the first
synthetic spectral analysis of the {\it IUE\/} archival spectra of KQ~Mon with
realistic optically thick, steady-state, viscous accretion-disk models with
vertical structure and high-gravity photosphere models. We find that the
photosphere of the white dwarf (WD) contributes very little FUV flux to the spectrum
and is overwhelmed by the accretion light of a steady disk. Disk models
corresponding to a white-dwarf mass of $\sim$$0.6\,M_\sun$, with an accretion
rate of order $10^{-9}\, M_\sun$/yr and disk inclinations between 
60$\degr$ and 75$\degr$,
yield distances from the normalization in the range of 144 to 165~pc. KQ~Mon is
discussed with respect to other nova-like variables. Its spectroscopic similarity to the FUV spectra of three definite SW~Sex stars suggests that it is likely a member of the SW~Sex class and lends support to the possibility that the WD is magnetic.

\end{abstract}

 
\section{Introduction}

Nova-like variables are a subset of cataclysmic variables (CVs), compact,
short-period binaries in which a late-type, Roche-lobe-filling main-sequence
dwarf transfers gas through an accretion disk onto a rotating, accretion-heated
white dwarf (WD). The spectra of most nova-like variables resemble those of classical
novae (CNe) that have settled back to quiescence. However, they have never had a
recorded CN outburst or any outburst. Hence, their evolutionary status is
unclear: they could be close to having their next CN explosion, or they may have
had an unrecorded explosion in the past, possibly hundreds or thousands of years
ago. Or, quite possibly, they may not even have CN outbursts at all, in which
case their WDs would be steadily increasing in mass to become the elusive
progenitors of Type~Ia supernovae. 

The variability of KQ~Mon was discovered by Hoffmeister (1943), who categorized
it as an irregular variable. Cameron \& Nassau (1956) classified the spectrum of
KQ~Mon as a carbon star based on a low-resolution objective-prism plate;
however, as noted by Stephenson (1989), this classification actually applies to
the nearby brighter and unrelated carbon star Case~432. The true nature of
KQ~Mon was revealed when Bond (1979), who had noted its neutral-to-blue color on
Palomar Sky Survey prints, obtained spectroscopy and high-speed photometry
showing it to be a CV. The spectra showed very shallow Balmer and \ion{He}{1}
lines in absorption, and the photometry showed low-amplitude flickering but no
obvious orbital modulation. These properties, combined with the lack of any known
large-amplitude outbursts or fadings, lead to the classification of KQ~Mon as a
nova-like variable.

Hoard et al.\ (2003) identified the object in the 2MASS survey and found that it
is a close visual triple star.
Schmidtobreick et al.\ (2005) determined the orbital period of KQ~Mon to be
3.08(4) hours using time-resolved spectroscopic data obtained over two nights at the Cerro Tololo Inter-American Observatory. 
They also suggested that KQ~Mon may be a member of the SW~Sex nova-like subclass
which are defined by a number of spectroscopic and photometric characteristics
including orbital periods between 3 and 4 hours, single-peaked emission lines
seen in high-inclination systems, high-excitation spectral features including
He~II ($\lambda$4686) emission, and strong Balmer emission on a blue continuum.
KQ~Mon's orbital period lies in the 3 to 4 hour period range, just above the
upper boundary of the CV period gap, where there is a pileup of SW~Sextantis
stars. It is classified as a probable SW~Sex star in the Big List of
SW~Sextantis Stars \footnote{See the Big List of SW~Sextantis Stars
at \url{http://www.dwhoard.com/biglist}}. We also note
that KQ~Mon is a cataloged X-ray source, 1RXS~J073120.7$-$102153 (Voges et al. 1999). 

In this paper, our objectives are to shed further light on the nature of KQ~Mon
by (1)~presenting new optical spectra obtained with the SMARTS telescope 33
years after the original KPNO IIDS spectra acquired by Bond (1979), and
(2)~derive fundamental parameters for KQ~Mon in the far ultraviolet by using
synthetic spectral, optically thick, steady-state accretion-disk models with
vertical structure and WD photosphere models, for the first time for this
system.The results of our model fitting should provide the first
estimates of the physical parameters for this nova-like variable and help
establish its classification  as a nova-like variable of the  UX~UMa subclass (nova-like systems that remain in a high state with no record of low states), a
member of the SW~Sex subclass type, or possibly an intermediate polar.  

\section{Optical Spectra}

Bond (1979) obtained intensified image-dissector scanner (IIDS) spectra with the
KPNO 2.1m telescope as part of a general spectroscopic survey of cataclysmic
variables that he carried out in the late 1970's with a resolution of about
7~\AA\ and covering 3500-5300~\AA. At the time, KQ~Mon had only been classified
as a slowly variable star (type ``L" in the GCVS). However, Bond (1979) had
noticed a neutral to  slightly blue color of KQ~Mon on the Palomar Sky Survey
prints, and thus added it to the IIDS observing program. On the basis of these
inital spectra of 1978 April 11, Bond noted shallow Balmer and He I (4471)
absorption similar to those in CD~$-42^\circ$14462 (V3885~Sgr), a well-known
UX~UMa nova-like star. KQ~Mon was observed with the IIDS again on 1978 Nov 25
and  1979 Oct 26, with little change noted relative to the first spectrum, both
with respect to their continuum slopes and absorption-line strengths.

We obtained an optical spectrum of KQ~Mon at a similar spectral resolution and
range of wavelength with the SMARTS 1.5m telescope and Ritchey-Chretien
spectrograph on 2011 Dec 21. In Fig.~1, we display a comparison of the continuum
slopes and absorption-line strengths between the SMARTS spectrum of 2011 and the
earlier spectra obtained with the IIDS. The 2011 spectrum appears to have a
shallower continuum slope and stronger Balmer absorption lines than the spectra
obtained in 1978-79. However, the continuum slope difference seen in the 2011
observation must be viewed with caution since a fairly narrow slit was used. The slit was oriented east-west and not at the parallactic angle, so that wavelength-dependent losses 
were indeed possible which would affect the blue end of the spectrum more than the red because the guider TV is red-sensitive. Because of these slit 
losses, and because only two spectrophotometric standards were observed (one at 
the beginning of the night, the second at the end), only the appearance of the 
spectrum should be taken into account, not the absolute flux or continuum slope.

For the IIDS spectra on the other hand, the continuum slopes are more reliable
since a larger aperture was used. The IIDS spectra in Fig.~1 agree quite well
with each other. While the flatter continuum slope in the 2011 spectrum taken at
face value might suggest a lower accretion rate in 2011, the increased
equivalent widths of the absorption features in 2011 (except H$\beta$ which is
almost completely filled in the 1978-79 and 2011 spectra) may imply a higher
accretion rate in 2011 (Warner 1995). This is especially evident in H$\delta$,
H$\epsilon$ and H8. 

\section{Far-Ultraviolet Spectra}

A total of eight spectra is available in the {\it International Ultraviolet
Explorer\/} ({\it IUE}) archive, five obtained with the short-wavelength prime
camera (SWP) and three obtained with the long-wavelength redundant (LWR),
between 1981 and 1983. These spectra were first analyzed by Sion and Guinan (1982) who used cruder disk models without vertical structure and black body fits. They also determined that the interstellar reddening was low due to the absence of an absorption dip at 2200~\AA. This result was
supported by la Dous (1991), who quoted an $E(B-V)$ value of 0.0 for KQ~Mon. 
 The inclination of the
system was unknown according to la Dous (1991) but probably less than
40$^{\circ}$,  based upon equivalent line widths of Si IV (1393, 1402), C IV
(1548, 1550), and Al III (1854, 1862).
 
We selected a well-exposed SWP spectrum for our synthetic model fitting,
SWP~15384, which was obtained at low resolution through the large aperture
starting on 1981-11-04 at 12:41:00 UT with an exposure time of 3000~s. As seen
in Fig.~2, the {\it IUE\/} SWP spectrum reveals the deep, high-ionization,
absorption features of C III (1175), N V (1240), Si II + O I (1300), C II
(1335), O V (1371), C IV (1550), He II (1640), N IV (1718), and possibly Al III
(1854). Even though the depths of the absorption lines suggest that the
accretion disk is being viewed at low inclination, it is unusual that P~Cygni
profiles are not seen, at least in C IV, as one sees in RW Sex, V3885 Sgr, IX
Vel and other UX~UMa systems, viewed at low inclination. It is also curious that
the centers of the absorption lines do not appear blue-shifted, which is a
measurable displacement at low {\it IUE\/} resolution in other UX~UMa systems
(Sion \& Guinan 1982).

\section{Synthetic Spectral Modeling}

We adopted model accretion disks from the optically thick disk model grid of
Wade \& Hubeny (1998). In these accretion-disk models, the innermost disk
radius, $R_{\rm in}$, is fixed at a fractional WD radius of  1.05.  The
outermost disk radius, $R_{\rm out}$, was chosen so that $T_{\rm eff}(R_{\rm
out}$) is near 10,000~K, since disk annuli beyond this point, which are cooler
zones with larger radii, would provide only a very small contribution to the
mid- and far-UV disk flux, particularly the SWP FUV bandpass. The mass-transfer
rate is assumed to be the same for all radii. For a given spectrum, we carry out
fits for every combination of $\dot M$, inclination, and WD mass in the Wade \&
Hubeny (1998) library. The values of $i$ are 18$^{\circ}$, 41$^{\circ}$,
60$^{\circ}$, 75$^{\circ}$, and 81$^{\circ}$. The range of accretion rates
covers $-10.5< \log(\dot{M})<-8.0 $ in steps of 0.5 in the log and five
different values of the WD mass, namely, 0.4, 0.55, 0.8, 1.0, and 1.2 solar
masses.  

Theoretical, high-gravity, photospheric spectra were computed by first using the
code TLUSTY (Hubeny 1988) to calculate the atmospheric structure and SYNSPEC
(Hubeny \& Lanz 1995) to construct synthetic spectra. We compiled a library of
photospheric spectra covering the temperature range from 15,000~K to 70,000~K in
increments of 1000~K, and a surface-gravity range of $\log g=7.0$ to 9.0, in
increments of 0.2.

After masking emission lines in the spectra, we determined separately for each
spectrum, the best-fitting WD-only models and the best-fitting disk-only models
using IUEFIT, a $\chi^{2}$ minimization routine. A $\chi^{2}$ value and a scale
factor were computed for each model fit. The scale factor, $S$, normalized to a
kiloparsec and solar radius, can be related to the WD radius $R$ through
$F_{\lambda}({\rm obs})=S\, H_{\lambda}({\rm model})$, where  $S=(4\pi
R^{2})/(d^{2})$ and $d$ is the distance to the source in parsecs. The details of
our $\chi^{2}$ (per degree of freedom) minimization fitting procedures are
discussed in detail in Sion et al.\ (1995). We take any reliable published
parameters like the WD mass, or orbital inclination only as an initial guess in
searching for the best-fitting accretion disk models. We search for the best fit
based upon (1) the minimum $\chi^{2}$ value achieved, (2) the goodness of fit of
the continuum slope, (3) the goodness of fit to the observed Ly$\alpha$ region
and (4) consistency of the scale factor-derived distance with the adopted
distance or distance constraint. In other words, the fitting solution may not
necessarily be the model with the lowest $\chi^{2}$ value but rather all other
available constraints are used such as, for example,                                                              constraints on the distance
or WD mass, and any reliable system parameters, if available. We utilize
absorption line profile fits, especially the Lyman alpha wings, but sometimes
even Si II, Si III, C III, C II, Si IV and Si II when these features are in
absorption and do not have an origin in a wind or corona. For the WD radii, we
use the mass-radius relation from the evolutionary model grid of Wood (1995) for
C-O cores. We also search for any statistically significant improvement in the
fitting by combining the FUV flux of a WD model and and a disk model, together
using a $\chi^{2}$ minimization routine called DISKFIT.  Once again, we find the
minimum $\chi^{2}$ value achieved for the combined models, and check the
combined model consistency with the observed continuum slope and Ly$\alpha$
region, and consistency of the scale factor-derived distance with the adopted
distance or distance constraints.  Out of the roughly 900 models using every
combination of $i$,  and M$_{wd}$, we try to isolate the best-fitting
accretion-disk model, WD model or combination of a disk and WD. 

The WD mass is unknown, as are the accretion rate,  orbital inclination,
and distance. We determined a solid lower 
limit to the distance using 2MASS {\it JHK\/} photometry  and the method of 
Knigge (2006). The firm lower limit to the distance is 100~pc, which provides 
an important constraint on the parameter range of our model fitting.

\section{Synthetic Spectral Fitting Results}
                                                                                                                                                                                                                                                                                                                                                                                      
Our procedure consisted of first trying photosphere models followed by optically thick accretion disk models. For the WD-only model fits, the WD solar composition photosphere models yielded
a minimum $\chi^{2}$ of 10.4802. This photosphere model gave a WD temperature of
21,000~K and a surface gravity of $\log g=8.0$, giving a distance to the system
of 333~pc. This model is shown in Fig.~3. We also tried a hotter photosphere
model (31,000~K), which was hot enough to have absorption lines of Si IV and C IV that fit the observed absorption features nicely, but the model continuum
longward of 1600~\AA\ was far too steep compared to the observed continuum.  

For the accretion-disk fitting, we tried disk models in the inclination range
18, 41, 60 and 75 degrees with the WD gravity fixed at $\log g = 8$. All of the
disk models, except two, gave marked shortfalls of flux longward of 1700~\AA.
Disk models corresponding to a WD mass of $0.55\, M_\sun$, accretion rates no
higher than  $10^{-9}\, M_\sun/yr$, and disk inclinations around 60$\degr$
yielded distances from the normalization in the range of 144 to 159~pc. The
best-fitting accretion-disk model with an inclination of 60$\degr$ to the
observation is displayed in Fig.~4. We note that a significantly  better fit is
achieved with an inclination of 75$\degr$, WD mass of $0.55\, M_\sun$, and an
accretion rate of $10^{-9} \,M_\sun$/yr. But the distance yielded by this fit is
only 90~pc, placing it below our adopted lower-limit distance of 100~pc from the
2MASS photometry. This fit is shown in Fig.~5.

We tried a combined fit (WD plus disk) to determine if a statistically
significant improved fit resulted compared with disks alone and with WD fits
alone, and to assess the relative flux contribution of disk and star to the
overall spectrum of the system.  The contribution of the photosphere turned out
to be minimal (about 3\%), being vastly outshone by the accretion light of the
disk. This combined fit is displayed in Fig.~6. Hence, the combination of the
photosphere with the disk model did not produce a statistically significant
improvement in the fit. Thus, it seems reasonable to state that the spectrum of
KQ~Mon is dominated by the accretion light of an optically thick disk.

We have summarized all of the model fitting results in Table~1.  The table
reveals that the steady accretion disk model fits, in  all cases, are markedly
superior to the model-photosphere fits. The accretion-disk models have $\chi^2$
values ranging between 4 and 6 while the  WD photosphere fits all have
significantly larger $\chi^2$ values, ranging between 10 and 17.

Moreover, if the FUV spectrum is due to a white dwarf, then all WD model fits with $T_{eff} >$ 31,000K were ruled out. The best WD fits to the observed FUV continuum were for temperatures too low to account for the strength of C IV (1550) and Si IV (1400). Thus, we can confidently rule out white dwarf photosphere models alone as the source of the FUV continuum.

Turning to the accretion disk fits in Table~1, we eliminated all of the disk model fits with WD mass, $M_{wd} < 0.4 M_{\odot}$ because masses this small are not seen among the CV population (except possibly for T Leonis; see Hamilton and Sion 2004). Given the firm lower limit to the distance of 100 pc that we derived from 2MASS photometry and the method of Knigge (2006), 
we eliminated all model fits with distances closer than 100 pc. The fact that eclipses have not been detected in KQ Mon allows us to eliminate model fits with i = 80 degrees or higher. A close examination of the plots corresponding to the remaining fits in 
Table~1 revealed that all of the disk model fits, except three, 
gave marked shortfalls of flux longward of 1700~\AA. 
Of the three remaining model fits, all have $M_{wd} = 0.55 M_{\odot}$, 
Two models have $\dot{M} = 10^{-9} M_{\odot}$/yr with i = 60 degrees while 
a third model has $\dot{M}= 3 \times 10^{-9} M_{\odot}$/yr with i = 75 degrees. All three models 
yielded scale factor-derived distances between 144~pc and 165~pc. From a formal error analysis on IUE spectra of comparable quality (Winter and Sion 2003), the estimated uncertainty in the accretion rates is a factor of two to three. Typical uncertainities for  photospheric temperatures of the
WDs, when exposed in the FUV, are $\pm$2000~K. 
 
\begin{table}[ht]
\centering
\caption{}{Synthetic Spectral Fitting Parameters}
\begin{tabular}{|c | c | c | c | c | c | c || c|}
\hline\hline
Spectrum & M$_{wd}$ & T$_{eff}$(WD) & $i$ (deg) & ${\dot M}$(M$_{\sun}/yr$) & $\chi^{2}$ & Scale & $d$(pc)\\ [0.5ex]
\hline    
\multicolumn{8}{|c|}{Disk Models} \\
\hline

swp15384&	0.35&	-&	60&	10$^{-8}$&	6.43&	0.077&	359\\	

swp15384&	0.35&	-&	60&	10$^{-9}$&	5.03&	1.259&	89\\

swp15384&	0.35&	-&	75&	10$^{-8}$&	4.89&	0.230&	208\\

swp15384&	0.35&	-&	81&	10$^{-8}$&	4.88&	0.651&	124\\

swp15384&	0.55&	-&  	41&	10$^{-9}$&	5.53&	0.228&	209\\

swp15384&	0.55&	-&	60&	10$^{-9}$&	4.45&	0.479&	144\\	

swp15384&	0.55&	-&	60&	10$^{-8.5}$&	6.26&	0.123&	284\\	

swp15384&	0.55&	-&	60&	10$^{-9.5}$&	4.99&	2.008&	71\\	

swp15384&	0.55&	-&	75&	10$^{-8.5}$&	4.78&	0.368& 	165\\	

swp15384&	0.55&	-&	75&	10$^{-9}$&	3.42&	1.568&	80\\	

swp15384&	0.8&	-&	60&	10$^{-9}$&	6.09&	0.208&	219\\	

swp16268&	0.35&	-&	75&	10$^{-8}$&	6.12&	0.190&	229\\	

swp16268&	0.35&	-&	81&	10$^{-8}$&	6.10&	0.538&	136\\	

swp16268&	0.55&	-&	60&	10$^{-9}$&	5.30&	0.396&	159\\	

swp16268&	0.55&	-&	75&	10$^{-8.5}$&	5.97&	0.304&	181\\	

swp16268&	0.55&	-&	81&	10$^{-8.5}$&    5.95&	0.859&	108\\

swp16268&	1.03&	-&	18&	10$^{-10}$&	6.63&	0.382&	162\\

swp16268&	1.21&	-&	18&	10$^{-10.5}$&	5.91&	0.745&	116\\
\hline
\multicolumn{8}{|c|}{WD Models} \\
\hline

swp15384&	0.5&	15000&	-&	-&	16.66&	0.409&	156\\

swp15384&	0.5&	21000&	-&	-&	10.48&	0.090&	333\\

swp15384&	0.5&	31000&	-&	-&	15.84&	0.017&	763\\

swp15384&	0.6&	21000&	-&	-&	11.36&	0.088&	337\\
\hline
\end{tabular}
\end{table}

\section{Conclusions}

Our synthetic spectral modeling of the FUV spectrum of KQ~Mon reveals that the
photosphere of the WD contributes little flux to the spectrum of the system, as
it is greatly overwhelmed by a luminous accretion disk. We find that the system
is being viewed at an inclination of no more than 60$\degr$, the mass of the WD
is $\sim$$0.6 \,M_\sun$, and the accretion rate is $\sim$$10^{-9}\, M_\sun$/yr
at a system distance between 144 and 159~pc. Since there is no record of a
previous outburst or of deep lower optical brightness states (like shown by
VY~Scl stars) with which to compare its current state, and since there is an
absence of low-excitation emission lines in its optical spectrum, it should be
classified as a member of the UX~UMa subclass of nova-like variables. However,
in UX~UMa systems viewed at low inclination, strong, variable, high-velocity
wind outflows are manifested in absorption features with large blueshifts and
prominent P~Cygni profile structure. Examples of these systems are RW Sex (Greenstein \& Oke 1982), IX
Vel (Sion 1985), V3885 Sgr (Linnell et al. 2009) , QU Car (Linnell et al. 2008), and RZ~Gru (Bisol et al.2012).

The question remains whether KQ~Mon is also an SW~Sex member. In addition to the
characteristics of the SW~Sextantis stars given in the introductory section,
they exhibit high-velocity emission $S$-waves with maximum blueshift near phase
$\sim$0.5, a delay of emission-line radial velocities relative to the  motion of
the WD, and central absorption dips in the emission lines around phase
$\sim$0.4--0.7 (Schmidtobreick et al.2005; Rodriguez-Gil, Schmidtobreick, \& Gaensicke 2007; Hoard et al.\
2003). The WDs in many if not all of these systems are suspected of being
magnetic (Rodriguez-Gil et al.\ 2007), although this hypothesis remains
unproven. Since SW~Sex stars as well as most of the UX~UMa systems are found
near the upper boundary of the two- to three-hour period gap, a much better
understanding of them is of critical importance to understanding CV evolution as
they enter the period gap (Rodriguez-Gil et al.\ 2007), if indeed they even do
enter the gap, since evolution across the gap has not yet been definitively
proven.

KQ~Mon is listed as a possible SW~Sex member in the Big List of SW~Sextantis
Stars primarily on the basis of variable, high-velocity wings seen
in optical spectra (Rodriguez-Gil 2005). Moreover, its orbital period ($P_{orb}
= 3.08$~hr) would place KQ~Mon at a position just above the period gap for
CVs. The nova-like variables in this period interval tend to
be strong candidates for being physical SW~Sex type stars (Rodriguez-Gil 2005). 

Curiously, the FUV absorption lines in KQ~Mon not only lack P~Cygni structure
but the line centers reveal only relatively small blue-shifts. The latter would
be the case if KQ~Mon is being viewed nearly face-on but the lack of P~Cygni
structure is puzzling. It is well known that polars and intermediate polars
typically do not exhibit wind outflow, with one possible exception
(H~0551$-$819; Mouchet et al.\ 1996). We speculate that it is possible that
SW~Sex stars contain WDs with magnetic fields well below the field strengths of
intermediate polars/polars but of sufficient strength to have a suppressive
affect on the wind-launching mechanism originating in the inner accretion disk.
According to this possible scenario, those UX~UMa nova-likes such as RW Sex, IX
Vel, V3885 Sgr, QU Car, and RZ Gru are non-magnetic and the wind launching
mechanism is unaffected.

We note that a comparison of the FUV spectra of KQ~Mon with other members of the
UX~UMa class may lend support to this possibility. We have identified three UX
UMa-type nova-likes, BP Lyn (Grauer et al. 1994, V795 Her (Shafter et al. 1990)  and LS Peg (Taylor et al. 1999), all with low inclinations ($<
60 \degr$) similar to KQ~Mon. It is far from clear that the line-formation
source region in KQ~Mon and the three aforementioned systems, lie in wind
outflows. All four objects have deep absorption features, variable line profiles
but no clear evidence of the strong wind outflow seen clearly in RW Sex, IX Vel,
V3885 Sgr, QU Car, and RZ Gru. Thus, the similarity of KQ~Mon's line spectrum to
the FUV spectra of the definite SW~Sex members, BP Lyn , V795 Her and LS Peg,
lends additional support for the classification of KQ~Mon as an SW~Sex member.

\medbreak

\section{Acknowledgements}

We are grateful to an anonymous referee whose comments helped to improve our
paper. It is a pleasure to acknowledge the support of this research by NSF grant
AST0807892 to Villanova University. Summer undergraduate research support was
also provided by the NASA-Delaware Space Grant Consortium. We thank F.~Walter
for scheduling the 2011 SMARTS 1.5m observation, and Rodrigo Hern\'andez for
obtaining the observation. HEB thanks the STScI Director's Discretionary
Research Fund for supporting STScI's membership in the SMARTS consortium.


\clearpage


\begin{figure}
\begin{center}
\includegraphics[width=6in]{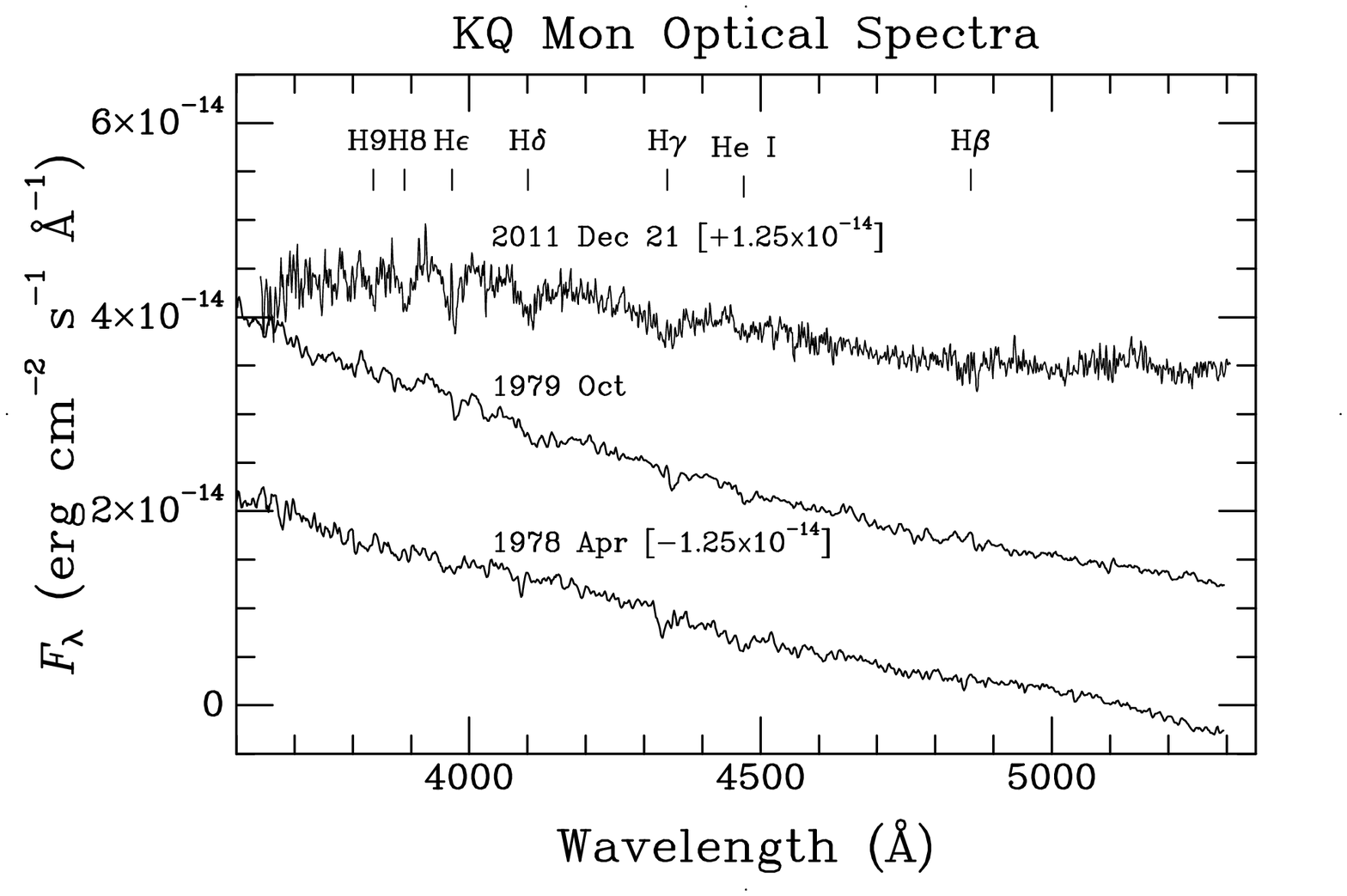}
\end{center}
\caption{Optical Spectra of KQ~Mon with the strongest absorption 
line features identified. The bottom two spectra were obtained 
in 1978-79 with the 2.1m KPNO telescope and the uppermost spectrum
 obtained in 2011 with the 1.5m SMARTS telescope. The plotted spectra 
have been offset from each other by the amounts indicated in the figure.  
The different slope in 2011 may not be reliable because of wavelength-dependent 
atmospheric-refraction losses. Note the stronger
Balmer absorption in 2011, and that H$\beta$ is filled by emission in all three
spectra.} 
\end{figure}

\begin{figure}[ht]
\begin{center}
\includegraphics[height=6.5in,angle=-90]{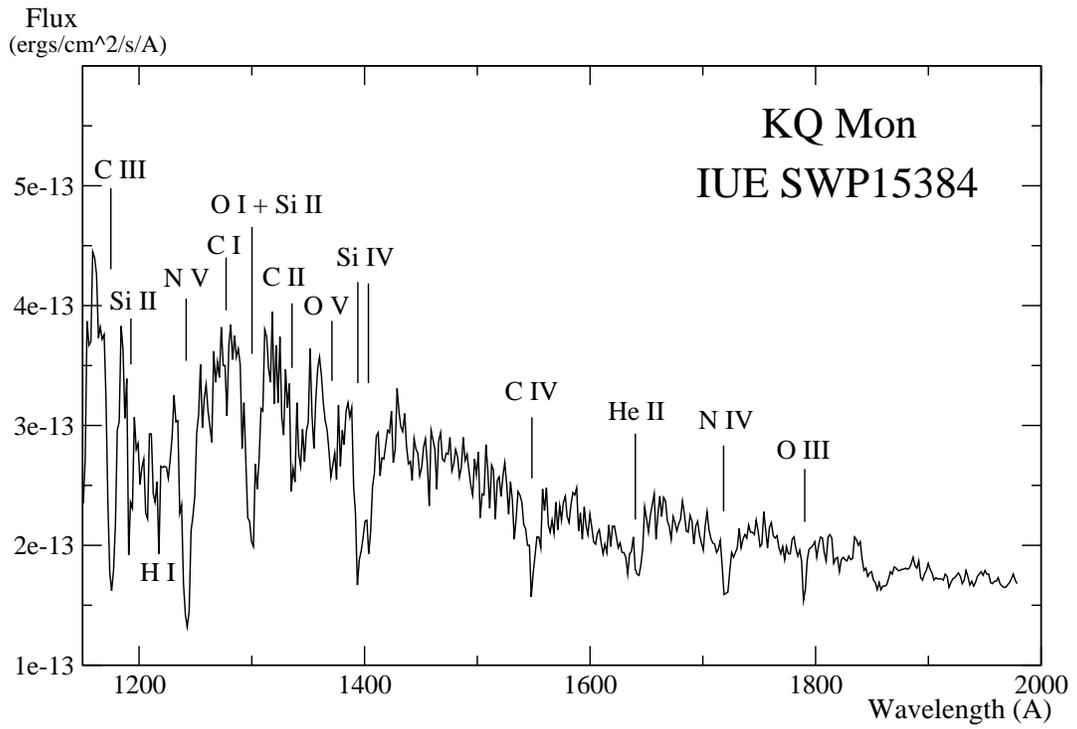}
\end{center}
\caption{The {\it IUE\/} spectrum SWP~15384 of KQ~Mon with the strongest
absorption line features identified. Note the lack of any P~Cygni line structure
or blue asymmetry.}
\end{figure}

\begin{figure}[ht]
\begin{center}
\includegraphics[height=6.5in,angle=-90]{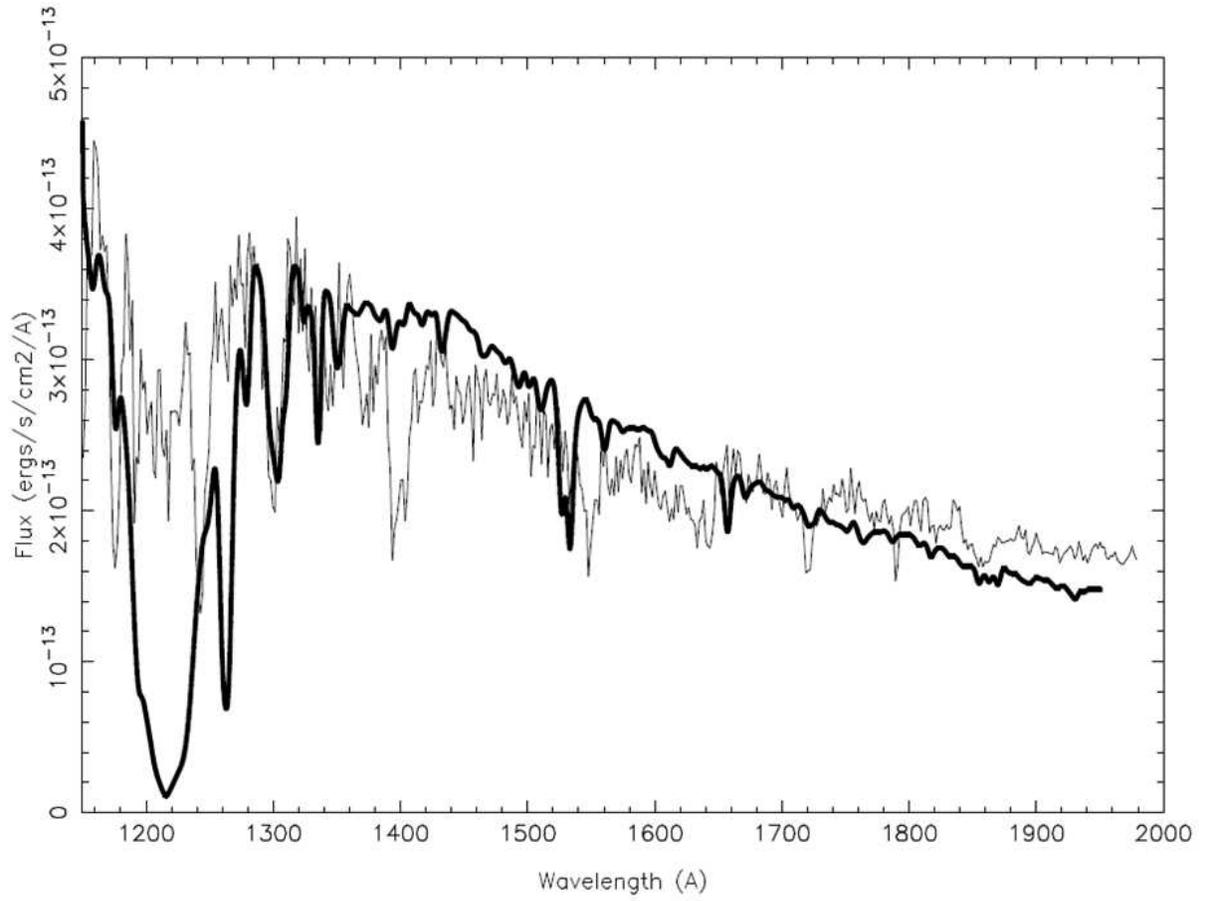}
\end{center}
\caption{A hot WD photosphere fit to SWP~15384 with $T_{\rm eff} = 21,000$~K,
$\log g = 8$
(see text for details).}
\end{figure}

\begin{figure}[ht]
\begin{center}
\includegraphics[height=6.5in,angle=-90]{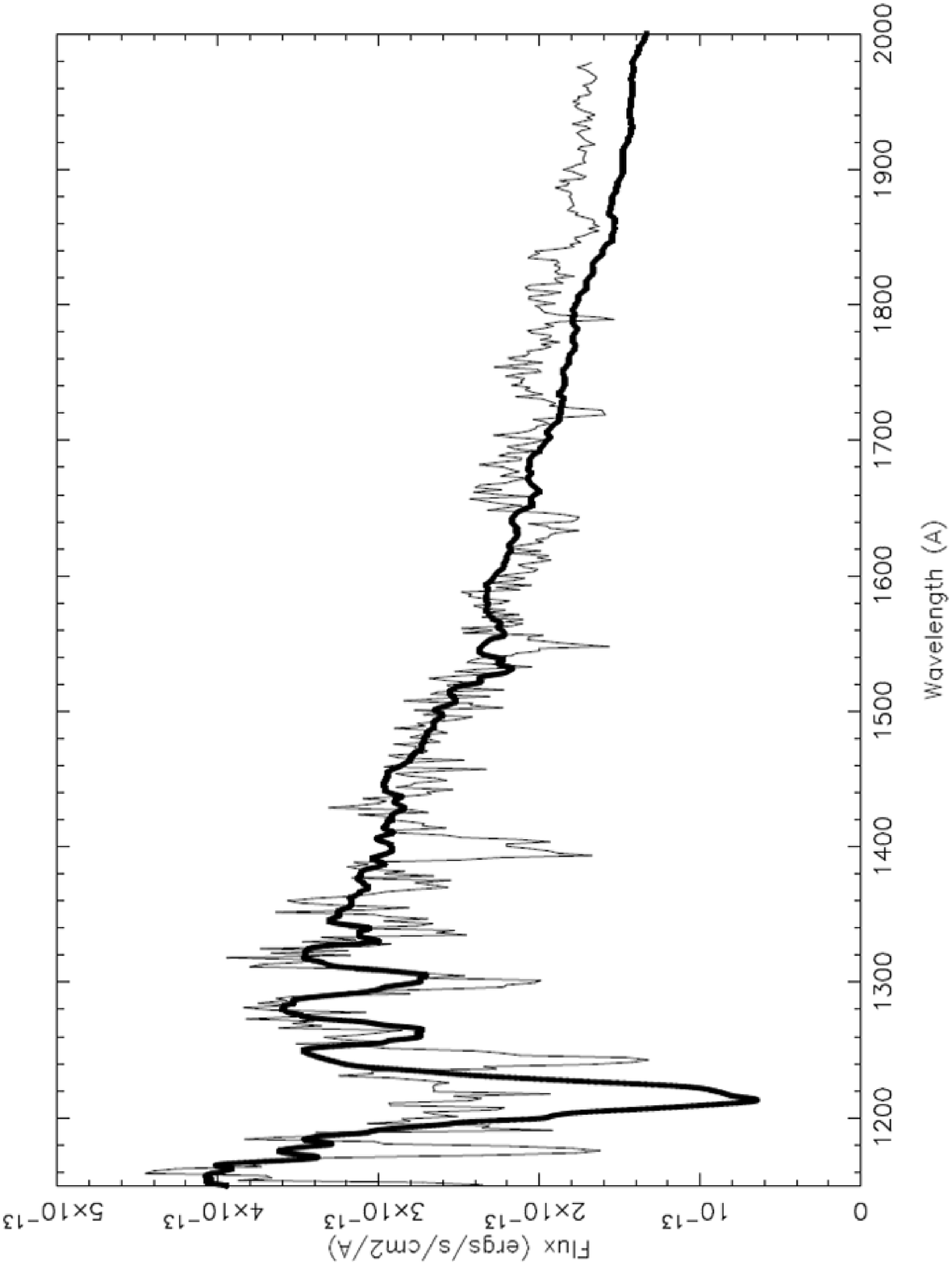}
\end{center}
\caption{The best fitting optically thick accretion disk model to SWP~15384, with
an inclination of 60$\degr$ (see text for details).}
\end{figure}

\begin{figure}[ht]
\begin{center}
\includegraphics[height=6.5in,angle=-90]{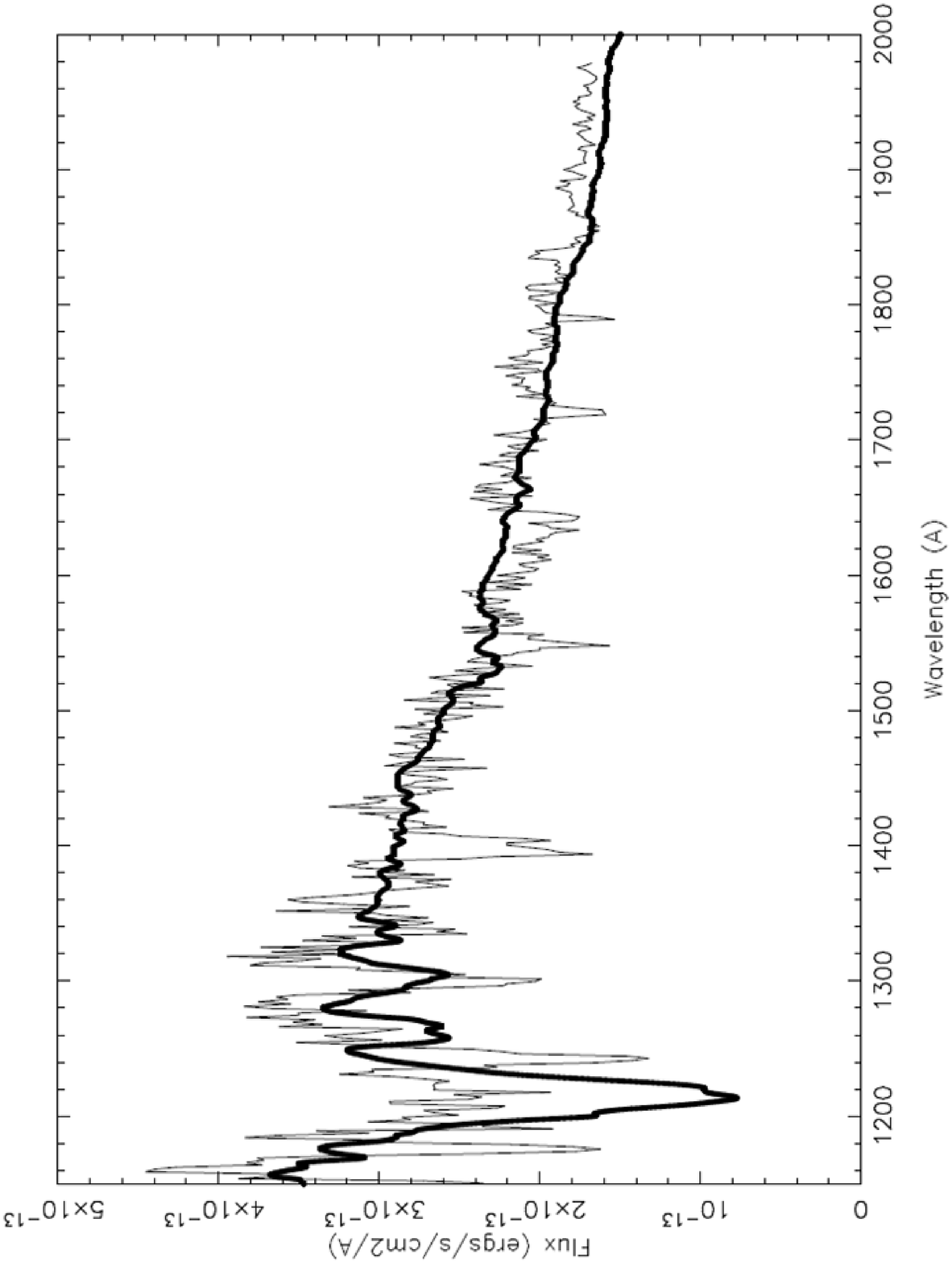}
\end{center}
\caption{The best fitting optically thick accretion disk model to SWP~15384 but
with an inclination of 75$\degr$ (see text for details).}
\end{figure}

\begin{figure}[ht]
\begin{center}
\includegraphics[height=6.5in,angle=-90]{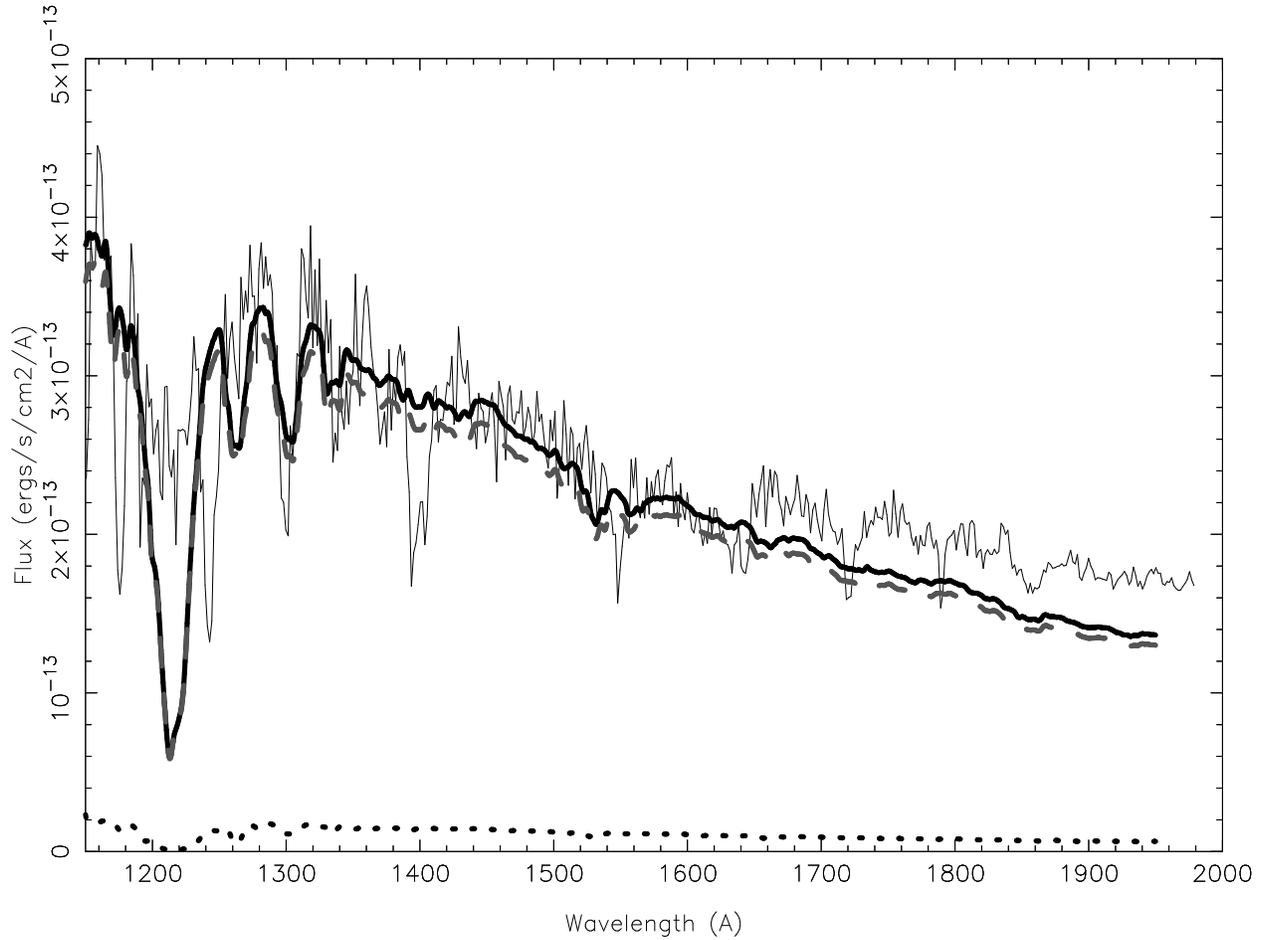}
\end{center}
\caption{Combination of a hot WD photosphere fit with $T_{\rm eff} = 21,000$~K, 
$\log g = 8$, and an optically thick, steady-state accretion disk to SWP~15384.
The accretion disk fit is shown as the dashed line, the WD model
by a dotted line, and the combined model fit by the solid line
(see text for details).}
\end{figure}

\end{document}